\documentclass[12pt,a4paper]{article}
\usepackage[left=1.2in, right=1.2in]{geometry}
\usepackage[numbers]{natbib}

\usepackage{times}
\usepackage{graphicx}
\usepackage{latexsym}
\usepackage{amssymb}
\usepackage{amsmath}
\usepackage[font={small,bf}]{caption}

\usepackage{paralist}

\usepackage{algorithm}
\usepackage{algpseudocode}
\algtext*{EndFor}
\algtext*{EndIf}

\DeclareMathOperator*{\argmin}{arg\,min}

\newcommand {\ignore} [1] {}

\newcommand{\cout}[1]{}

\usepackage[]{hyperref}
\hypersetup{pdftex,colorlinks=true,allcolors=blue}

\begin{document}

%

\title{ERA: A Framework for Economic Resource Allocation for the Cloud}
%
%
%
%
%

%

\author{
%
%
Moshe Babaioff\thanks{Microsoft Research, moshe,ishai@microsoft.com}
\and
Yishay Mansour\thanks{TAU and Microsoft Research, mansour@microsoft.com} 
\and
Noam Nisan\thanks{HUJI and Microsoft Research, noam,gali.noti@cs.huji.ac.il} 
\and
Gali Noti\footnotemark[3]
\and
Carlo Curino\thanks{Microsoft, ccurino,narg@microsoft.com} 
\and
Nar Ganapathy\footnotemark[4]
\and
Ishai Menache\footnotemark[1]
\and
Omer Reingold\thanks{Stanford University, reingold@stanford.edu}
\and
Moshe Tennenholtz\thanks{Technion and Microsoft, moshet@ie.technion.ac.il, ereztimn@cs.technion.ac.il} 
\and
\and
Erez Timnat\footnotemark[6]
}

\newcommand{\cch}[1]{\textcolor{black}{#1}}

\maketitle
\begin{abstract}

Cloud computing has reached significant maturity from a systems perspective, 
but currently deployed solutions rely on rather basic economics mechanisms that yield suboptimal allocation of the costly hardware resources.
In this paper we present Economic Resource Allocation (ERA), a complete framework for scheduling and pricing cloud resources,
aimed at increasing the efficiency of cloud resources usage by allocating resources 
according to economic principles. 
The ERA architecture carefully abstracts the underlying cloud infrastructure, 
enabling the development of scheduling and pricing algorithms 
independently of the concrete lower-level cloud infrastructure and independently of its concerns. 
Specifically, ERA is designed as a flexible layer that can sit on top of any cloud system and 
interfaces 
with both the cloud resource manager and 
with the users who {\em reserve} resources to run their jobs. 
The jobs are scheduled based on prices that are {\em dynamically} calculated according to the predicted demand. 
Additionally, ERA provides a key internal API to {\em pluggable} algorithmic modules that include scheduling, pricing and demand prediction. 
We provide a proof-of-concept software and demonstrate the effectiveness of the architecture by testing 
ERA over both public and private cloud systems -- Azure Batch of Microsoft and Hadoop/YARN. 
A broader intent of our work is to foster collaborations between economics and system communities. 
To that end, we have developed a simulation platform via which economics and
system experts can test their algorithmic implementations.

\end{abstract}

\section{Introduction}

Cloud computing, in its private or public incarnations, is commonplace in industry as a
paradigm to achieve high return on investments (ROI) by sharing massive computing
infrastructures that are costly to build and operate \cite{berkeley_cloud,greenberg2008cost}.
Effective sharing pivots around two key ingredients: 1) a {\em system infrastructure} that can
securely and efficiently multiplex a shared set of hardware resources among several tenants, and 2) {\em economic mechanisms} to arbitrate between conflicting resource demands from multiple tenants.

State-of-the-art cloud offerings provide solutions to both, but with varying degrees of sophistication.
The system challenge has been subject to extensive research focusing on space and time multiplexing. Space multiplexing consists of sharing servers among tenants, while securing them via virtual machine \cite{vmware,hyperv,xen} and container technologies \cite{docker,cgroups}. Time multiplexing comprises a collection of techniques and systems that schedule tasks over time. The focus ranges from strict support of Service Level Objectives (SLOs) \cite{curino2014rayon,morpheus,tetrisched,jockey} to maximization of cluster utilization \cite{drf,delay-scheduling,mercury,yaq,Sparrow-tr,tetris,apollo}.
Many of these advances are already deployed solutions in the public cloud \cite{hdinsight,elasticmapreduce} and the private cloud
\cite{hadoop,yarn,mesos,borg,apollo}. This indicates a good degree of maturity in how the system challenge is tackled in cloud settings.

On the other hand, while the economics challenge has received some attention in recent research 
(see, e.g., \cite{buyya2012pricing,yaniv1, azar2015truthful,song2012optimal,kilcioglu2016competition, menache2011socially} and references therein),  
the underlying principles have not yet been translated into equally capable solutions deployed in practice.

\subsection{The Economic Challenge and ERA's Approach}
In current cloud environments, resource allocation is governed by very basic economics mechanisms.
The first type of mechanism (common in private clouds \cite{yarn,apollo,mesos,borg}) uses fixed pre-paid guaranteed quotas.
The second type (common in public clouds \cite{hdinsight,elasticmapreduce}) uses on-demand unit prices: the users are charged real money\footnote{Or, within
a company, fiat money.} per unit of resource used.  In most cases these are fixed prices, with the notable exception
of Amazon's spot instances that use dynamically changing prices.\footnote{Although, based on independent analysis, even these may not truly leverage market mechanisms to determine prices \cite{agmon2013deconstructing}.} Spot-instance offerings, however, do not provide guaranteed service, as the instances might be evicted if the user bid is too low. Hence, utilizing spot instances requires special attention from the user when determining his bid, and might not be suitable for high-priority production jobs \cite{menache2014demand,song2012optimal,abhishek2012fixed}. 
The fundamental problem is finding a pricing and a scheduling scheme that will result in highly desired outcome, that of high efficiency.   

\noindent {\bf Efficiency: }
From an economics point of view, the most fundamental goal for a cloud system is to maximize the economic {\em efficiency}, that is, to maximize the total value that all users get from the system.
For example, whenever two users have conflicting demands, \cch{the one with the lowest cost for switching to an alternative (running at a different time/place or not running at all) should be the one switching. The resources would thus be allocated to the user with ``highest marginal value.''}
The optimal-allocation benchmark for a given cloud 
is that 
of an omniscient scheduler who has access to the complete information of all cloud users -- including
their internal costs and alternative options -- and decides what resources to allocate to whom in a way that maximizes the
efficiency goals of the owner of the cloud. 
Let us stress:
to get a meaningful measure of efficiency we must count the {\em value obtained} rather than the {\em resources used}, and we should aim to maximize this value-based notion of efficiency.\footnote{Another important goal, of course, is revenue, but we note that the potential revenue is limited by the value created, so high efficiency is needed for high revenue. Moreover, since there is competition between cloud providers, these providers generally aim to increase short-term efficiency as this is likely to have positive long-term revenue effects.
The issue of increasing revenue is usually attacked under the ``Platform as a Service'' (PaaS) strategy of providing higher-level services. This
is essentially orthogonal to allocation efficiency and is beyond the scope of the present paper.}

\noindent {\bf Limitations of current solutions: } 
With this value-based notion of efficiency in mind, let us evaluate commonly deployed pricing mechanisms. Private cloud frameworks  \cite{yarn,apollo,mesos,borg} typically resort to pre-paid guaranteed quotas. 
The main problem with this option is that, formally, it provides no real sharing of common resources: to guarantee that every user 
always has his guaranteed pre-paid resources available, the cloud system must actually hold sufficient resources to satisfy the sum of all promised capacities, even though only a
fraction will likely be used at any given time. 
\cch{Mechanisms such as work-preserving fair-queueing \cite{drf} are typically designed
to increase utilization \cite{yarn,mesos}, but they do not fundamentally
change the equation for value-based efficiency, as the resources offered
above the user quota are typically distributed at no cost and
with no guarantees.} 
Furthermore, lump-sum pre-payment implies that the users' marginal cost for using their guaranteed resources is essentially zero, and so they will tend to use their capacity for ``non-useful'' 
jobs whenever they do not fill their capacity with
``useful'' jobs.  This often results in cloud systems that seem to be operating at full capacity from every engineering point of view, but are really working at very ``low capacity'' from an economics point of view, as most of the time, most of the jobs have very low value.

\cch{On the other hand, public cloud offerings \cite{hdinsight,elasticmapreduce} typically employ on-demand unit-pricing schemes.} The issue with this solution is that the availability of resources cannot be guaranteed in advance.  Typically the demand is quite spiky, with short
periods of peak demand interspersed within much longer periods of low demand.  Such spiky demand is also typical of many other
types of shared infrastructure
such as computer network bandwidth, electricity, or ski resorts.  In all these cases the provider of the shared resource faces a dilemma between
extremely expensive over-provisioning of capacity and giving up on guaranteed service at peak times.  In the case of cloud systems,
for jobs that are important enough, users cannot take the risk of their jobs not being able to run when needed, and thus ``on-demand''
pricing is only appealing to low-value or highly time-flexible jobs, while most important ``production'' jobs with little time flexibility resort to buying long-term
guaranteed access to resources. 
While flexible unit prices (such as Amazon's spot instances) may have an advantage over fixed ones as they can better smooth demand over time,
as highlighted above, they 
only get an opportunity to do so for the typically low-value ``non-production'' jobs.

\noindent {\bf The ERA approach: } The pricing model that we present in ERA enables sharing of resources and smoothing of demand even for high-value production jobs.
This is done using
the well-known notion of  {\em reservations}, commonly used for many types of resources such as hotel rooms or super-computer time as well as in a few cloud systems
\cite{curino2014rayon,morpheus,tetrisched},
but in a {\em flexible way in terms of both pricing and scheduling.}
We focus on the economic challenges of scheduling and pricing batch style computations with completion-time SLOs (deadlines) on a shared cloud infrastructure.
The basic model presented to the user is that of {\em resource reservation}.  At ``reservation
time,'' the user's program specifies its reservation request.  The basic form of such a request is:\footnote{The general form of requests
is given by ERA's
``bidding description language'' (see Section \ref{sec:era-user}) that allows specifying multiple resources, variable ``shapes'' of use across time,
and combinations thereof.}

\vspace{2mm}
{\em {\bf Basic Reservation:} ``I need 100 containers (with 6GB and 2cores each) for 5 hours, some time between 6am and 6pm today, and am willing to pay up to \$100 for it.''}
\vspace{2mm}

This class of workloads is very prominent -- much of ``big data'' falls under this category \cite{jockey,tetrisched,morpheus} -- and it provides us with a unique opportunity for economic investigation. While state-of-the-art solutions provide effective system-level mechanisms for sharing resources, they rely on users' goodwill to truthfully declare their resource needs and deadlines to the system.
By dynamically manipulating the price of resources, ERA provides users with incentives to expose to the system as much flexibility as possible.
The ERA mechanism ensures that the final price paid by the user is the {\em lowest possible price} the system can provide for granting the request.
The more flexibility a user exposes, 
the better 
the user's chances of getting 
a good price. 
If this minimal price exceeds the stated maximal willingness to pay, then the
request is denied.\footnote{Alternatively, the ERA mechanism may present this minimal price as a ``quote'' to the user, who may decide whether to accept or reject it.}
Once a reservation request is accepted, the payment is fixed at reservation time, and
the user is assured 
that the reserved resources will be
available to him within the requested window of time.
The guarantee is to satisfy the request rather than provide a promise of specific resources at specific times.
For more details regarding the model presented to the user see Section \ref{sec:era-user}. 

\subsection{An Overview of ERA}

A key part of the challenge of devising good allocation schemes for cloud resources is their multi-faceted nature: while our goals are
in terms of economic high-level business considerations, implementation must be directly carried out at the 
computer systems-engineering level.  These two extreme points of view must be connected using clever algorithms and
implemented using appropriate software engineering.  Indeed, in the literature regarding cloud systems, one can see many papers
that deal with each one of these aspects -- ``systems'' papers as well as theoretical papers on scheduling and pricing jobs in cloud systems -- as cited above.
Unfortunately, these different threads in the literature are often disconnected from each other and they are not easy to combine to get an overall solution.
We believe that one key challenge at this point is to provide a common abstraction that encompasses all these considerations. We call this the {\em architectural challenge}.

Our answer to this challenge is the {\em ERA system} (Economic Resource Allocation).
The ERA system is designed as an intermediate layer between the cloud users and the underlying cloud infrastructure.
It provides a single model that encompasses all the very different key issues underlying cloud systems: economic, algorithmic, systems-level, and human-interface ones.
It is designed to integrate economics insights practically in real-world system infrastructures,
by guiding the resource allocation decisions of a cloud system in an economically principled way.\footnote{Not all resources in the cloud have to be managed via ERA. It is also possible that the cloud will let ERA manage only a subset of the resources (allowing the system to be incrementally tested), or will have several instances of ERA to manage different subsets of the resources.}
This is achieved by means of a key architectural abstraction: {\em the ERA Cloud Interface}, which hides many of the details of the resource management infrastructure,
allowing ERA to tackle the economic challenge almost independently of the underlying cloud infrastructure. 
ERA satisfies three key design goals: 1) it provides a crisp theoretical abstraction that enables more formal studies; 2) it is a practical end-to-end software system; and 3) it is designed for extensibility, where all the algorithms are by design easy to evolve or experiment with. 

\begin{figure}
	\centerline{\includegraphics[scale=0.4]{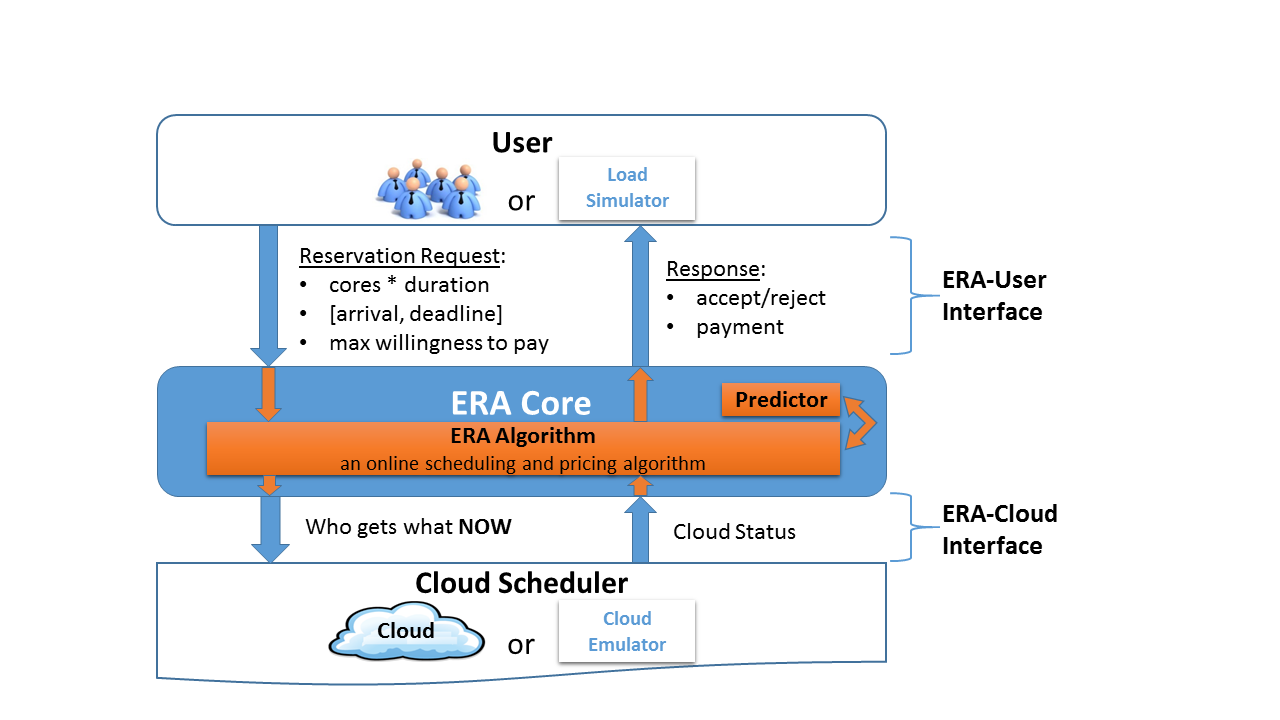}}
	\caption{ERA Architecture. The ERA system is designed as an intermediate layer between the users and the underlying cloud infrastructure. The same actual core code is also interfaced with the simulator components. \label{fig:architecture1} }
\end{figure}

\noindent {\bf ERA's key APIs: } ERA has two main outward-facing APIs as well as a key internal API.
Figure \ref{fig:architecture1} gives a high-level schematic of the architecture.  
The first external API faces the users and
provides them with the economic reservation model of cloud services described above.
The second external API faces the
low-level cloud resource manager. It provides
a separation of concerns that frees the underlying cloud system from any time-dependent scheduling or from
any pricing concerns, and frees the ERA system from the burden of assigning specific processors to specific tasks in a reasonable resource-locality 
way, or from the low-level mechanics of firing up processes or swapping them out.
See more details in Section \ref{sec:era-arch}.

Finally, the internal API is to pluggable algorithmic scheduling, pricing, and prediction modules.
Our basic scheduling and pricing algorithm dynamically computes future resource prices based on supply and demand, where the demand includes
both resources that are already committed to and predicted future requests, and schedules and prices the current request at the ``cheapest'' possibility.
Our basic prediction model uses traces of previous runs to estimate future demand.  The flexible algorithmic API then
allows for future algorithmic, learning, 
and economic optimizations.
The internal interfaces as well as our basic algorithmic implementations are described in Section \ref{sec:internal}.

\begin{figure}
	\centerline{\includegraphics[scale=0.40]{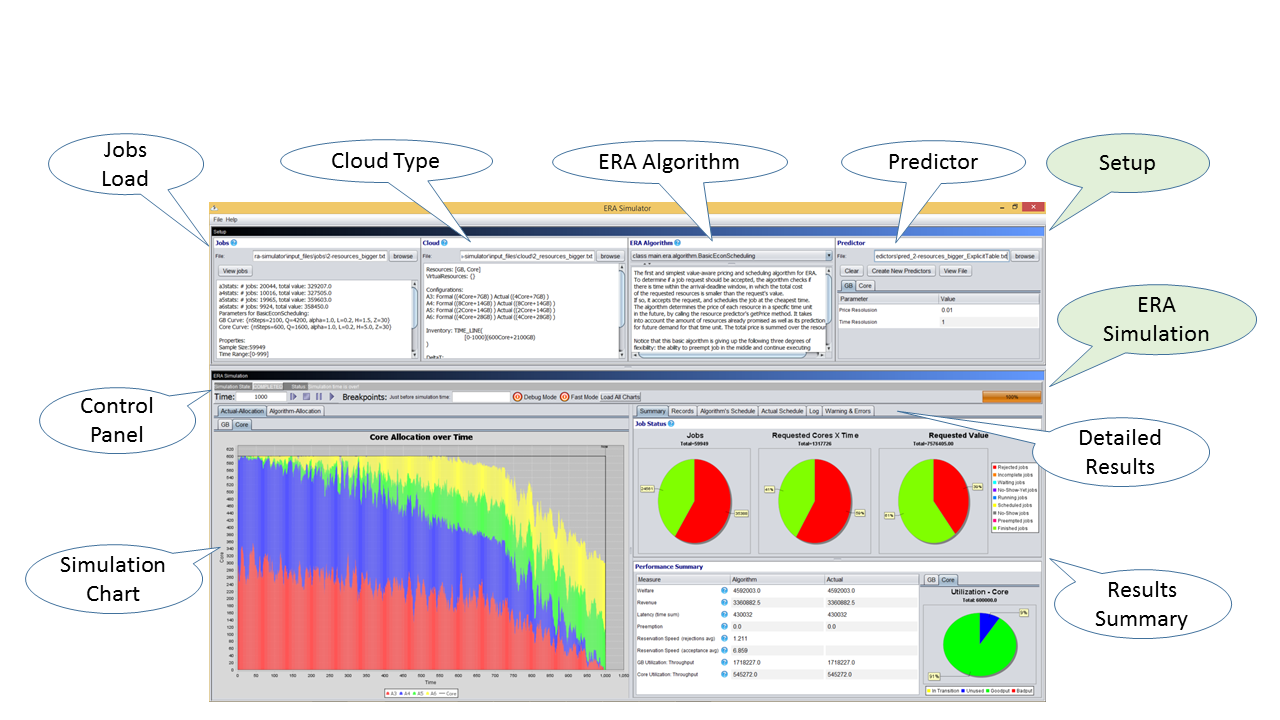}}
	\caption{ERA Simulator Screenshot \label{fig:simulator} }
\end{figure}

Our goal in defining this abstraction is more ambitious than mere good software engineering in our system.
As part of the goal of fostering a convergence between system and economic considerations, we have also built a flexible cloud simulation framework.
The simulator provides
an evaluation of key metrics, both ``system ones'' such as loads or latency, as well as ``economic ones'' such as ``welfare'' or revenue, as well as
provides a visualization of the results
(see screenshot in Figure \ref{fig:simulator}).  The simulator was designed to provide a convenient tool both for the cloud system's
manager who is interested in evaluating ERA's performance as a step toward integration and for
researchers who develop new algorithms for ERA and are interested in experimenting with their implementation without the need to run a large cluster.
As is illustrated in Figure \ref{fig:architecture1}, the same core code that receives actual user requests and runs over the underlying cloud resource manager may be
connected instead to the simulator so as to test it under variable loads and alternative cloud models.
Comparing the results from our simulator and physical cluster runs, we find the simulator 
to be faithful  (Section \ref{sec:simulator}).


\begin{sloppypar}
The ERA system is implemented in Java, and an alternative implementation (of a subset of ERA) in C\# was also done.
We have performed extensive runs of ERA within the simulator as well as 
proof-of-concept runs with two prominent resource managers in the public and private clouds: 
the full system was interfaced with Hadoop/YARN \cite{yarn} 
and the C\# version of the code was interfaced and tested with Microsoft's Azure Batch\footnote{Azure Batch is a cloud-scale job-scheduling and compute management service.
https://azure.microsoft.com/en-us/services/batch/} simulator \cite{azurebatch}. 
These runs 
show that the ERA algorithms succeed in increasing the efficiency of cloud usage, and that ERA can be successfully integrated with real cloud systems. Additionally, we show that the ERA simulator gives a good approximation to the actual run on a cloud system and thus can be a useful tool for developing and testing new algorithms.
In Section \ref{sec:simulator} we present the results of a few of these runs.
\end{sloppypar}

\vspace{0.1in}
\noindent
{\bf Contributions: }
In summary, we present ERA, a reservation system for pricing and scheduling jobs with completion-time SLOs. ERA makes the following contributions:
\begin{enumerate}
	\item We propose an abstraction and a system architecture that allows us to tackle the economic challenge orthogonally to the underlying cloud infrastructure.
	\item We devise algorithms for scheduling and pricing batch jobs with SLOs, 
	and for predicting resource demands.
	\item We design a faithful cloud simulator via which economics and
system experts can study and test their algorithmic implementations.
	\item We integrate ERA with two cloud infrastructures and demonstrate its effectiveness experimentally.
\end{enumerate}

\section{The ERA Model and Architecture}\label{sec:model}

\subsection{The Bidding Reservation Model with Dynamic Prices}

ERA is designed to handle a set of computational resources of a cloud system, such as cores and memory, with the goal of allocating these resources to users efficiently. 
The basic idea is that a user that needs to run a job at some future point in time can make a reservation for the required resources and, once 
the reservation is 
accepted, these 
resources 
are then guaranteed (insofar as physically possible) to be available at the reserved time. The guarantee of availability of reserved resources allows high-value jobs to use cloud just like in the pre-paid guaranteed quotas 
model, but without the need to buy the whole capacity (for all times), which thus also allows for time sharing of resources, which increases efficiency. 

The price for these resources is quoted at reservation time and is dynamically computed according to (algorithmically estimated) demand and the changing supply.  
More user flexibility in timing is automatically rewarded by lower prices.  The basic idea is that these dynamic prices will regulate demand, achieving a better utilization of cloud resources.
This mechanism also ensures that at peak times -- where demand can simply not be met -- the most ``valuable'' jobs are the ones that will be run rather than arbitrary ones.  
ERA uses a simple bidding model in which the user attaches to each of his job requests a monetary value specifying the maximal amount he is willing to pay for running the job. 
The amount of value lost for jobs that cannot be accommodated at these peak times serves as a quantification of the value that will be gained by buying additional cloud resources, and is an important input to the cloud provider's purchasing decision process.

\subsection{The Cloud Model} \label{sec:cloud_model}

The cloud in the ERA framework is modeled as an entity that sells multiple resources, bundled in configurations, and 
the 
capacity of these resources 
may change 
over time. 
The configurations, as well as the new concept of ``virtual resources,'' are designed to represent constraints that the cloud is facing, such as packing constraints. 
Specifically, the cloud is defined by: 
(1) a set of formal resources for sale 
(e.g., core or GB). 
We also allow for capturing additional constraints of the underlying infrastructure by using a notion of ``virtual resources''; 
(2) a set of resource configurations: 
each configuration is defined by a bundle of formal resources (e.g., ``ConfA'' equals 4 cores and 7 GB),\footnote{These configurations are preset, but notice that ERA's cloud model also supports the flexibility that each job can pick its own bundle of resources (as in YARN) by defining configurations of the basic formal resources (e.g., ``ConfCore'' equals a single core).} 
and is also associated with a bundle of actual resources  
that reflects the average amount the system needs in order to supply the configuration. 
The actual resources will typically be larger than the formal resources. The gap is supposed to model the overhead the cloud incurs when trying to allocate the formal amount of resources within a complex system. The actual resources can be composed of formal as well as virtual resources; 
(3) inventory: the amount of resources (formal and virtual) over time;
(4) time definitions of the cloud (e.g., the precision of time units that the cloud considers).

\subsection{The ERA Architecture} \label{sec:era-arch}

The ERA system is designed as a smart layer that lies between the user and the cloud scheduler, as shown in Figure \ref{fig:architecture1}. The system receives a stream of job reservation requests for resources arriving online from users. 
Each request describes the resources the user wishes to reserve and the time frame in which these resources are desired, as well as an economic value specifying the maximal price the user is willing to pay for these resources. 
ERA grants a subset of these requests 
with the aim of maximizing 
total value (and/or revenue). 
The interface 
with 
these user requests is described in Section \ref{sec:era-user}.

\begin{sloppypar}
ERA interfaces with the cloud scheduler to make sure that the reservations that were granted actually get the resources they were promised. ERA instructs the cloud how it should allocate its resources to jobs, and the cloud should be able to follow ERA's instructions and (optionally) to provide updates about its internal state (e.g., the capacity of available resources),  
allowing ERA to re-plan and optimize. 
ERA's interface 
with 
the cloud scheduler is described in Section \ref{sec:era-cloud}. 
\end{sloppypar}

The architecture encapsulates the logic of the scheduling and pricing in the \textit{algorithm} module. 
The algorithms use the {\em prediction} module to compute prices dynamically based on the anticipated demand and supply. This architecture gives the ability to change between algorithms and to apply different learning methods. ERA's internal interface 
with 
the algorithmic components is described in Section \ref{sec:internal}. 


\subsubsection{ERA-User Interface} \label{sec:era-user}

The ERA-User interface handles a stream of reservation requests that arrive online from users, and determines which request is accepted and at which price. 

\paragraph{The Bidding Description Language}
Each reservation request bids for resources according to ERA's bidding description language -- an extension of the reservation definition language formally defined in \cite{curino2014rayon}. 
The bid is composed of a list of \textit{resource requests} and a maximum willingness to pay for the whole list.
Each resource request specifies the configurations of resources that are requested, the length of time for which these are needed, and a time window $[arrival, deadline)$. All the resources must be allocated after the arrival time (included) and before the deadline (excluded). 
For example, a resource request may ask for a bundle of 3 units of ConfA and 2 units of ConfB, for a duration of 2 hours, sometime between 6AM and 6PM today. Each configuration is composed of one or more resources, as described in Section \ref{sec:cloud_model}.

By supporting a list of resource requests, ERA allows the description of more complex jobs, including the ability to break each request down to the basic resource units allowing for MapReduce  kinds of jobs, or to specify order on the requests to some degree.
The current ERA algorithms accept a job only if all of the resource requests in the list can be supplied; i.e., they apply the AND operator between resource requests. 
More sophisticated versions may allow more complex and rich bidding descriptions, e.g., support of other operators or recursive bids. 
For clarity of presentation, in this paper we present ERA in the simple case, where the reservation request is a single resource request, and there is only a single resource rather than configurations of multiple resources.

\paragraph{The \textit{makeReservation} method}
The interface with the user reservation requests 
is composed of the single ``\textit{makeReservation}'' method, which handles a job reservation request that is sent by some user. 
Each reservation request can either be accepted and priced or declined by ERA.

The basic input parameters to this method are the job's bid and the identifier of the job.
The bid encapsulates a list of resource requests along with the maximum price that the user is willing to pay in order to get this request (as described above). The output is an acceptance or rejection of the request, and the price that the user will be charged for fulfilling his request in case of acceptance.\footnote{Alternatively, the system may allow determining the payment after running is completed (depending on the system load at that time), or may allow flexible payments that take into account both the amount of resources reserved and the amount of resources actually used.}

The main effect of accepting a job request is that 
the user 
is guaranteed 
to be given the desired amount of resources sometime within the desired time window. 
An accepted job must be ready to use all requested resources starting at the beginning of the requested window, and the request is considered fulfilled as long as ERA provides the requested resources within the time window.

\subsubsection{ERA-Cloud Interface} \label{sec:era-cloud}

The interface between ERA and the cloud-scheduler is 
composed of two main methods that allow the cloud to get information about the allocation of resources it should apply at the current time, and to provide ERA with feedback regarding the actual execution of jobs and changes in the cloud resources.

\paragraph{The \textit{getCurrentAllocation} method} 
This is the main interface with the actual cloud scheduler. 
The cloud should repeatedly call this method (quite often, say, every few seconds) and ask ERA for the current allocations to be made.\footnote{For performance, it is also possible to replace this query with an event-driven scheme in which ERA pushes an event to the cloud scheduler when the allocations change. } 
The method returns an \textit{allocation}, which is 
the  list of jobs that should be instantaneously allocated resources and the resources that should be allocated to them. 
In the simple case of a single resource, it is a list of ``job J should now be getting W resources.'' 
The actual cloud infrastructure should update the resources that it currently allocates to all jobs to fit the results of the current allocation returned by this query. This new allocation remains in effect until a future query returns a different allocation. 
It is the responsibility of the underlying cloud scheduling system to query ERA often enough, and to put these new allocations into effect ASAP, so that any changes are effected with reasonably small delay. 
The main responsibility of the ERA system is to ensure that the sequence of answers to this query reflects a plan that can accommodate all accepted reservation requests.

The main architectural aspect of this query is to make the interface between ERA and the cloud system narrow, such that it completely hides the plan 
ERA has for future allocation. 
It is assumed that the cloud has no information on the total requirements of the jobs, and follows ERA as accurately as possible.

\paragraph{The \textit{update} method (optional usage)} 
The cloud may use this method to periodically update ERA with its actual state. Using this method is important 
since the way resources are actually used in real time may be different from what was planned for. 
For example, 
some processors may fail or be taken offline. Most importantly, it is expected that most jobs will use significantly less resources than what they reserved (since by trying to ensure that they have enough resources to usually complete execution, they will probably reserve more than they actually use). The ERA system should take this into account and perhaps re-plan.

The simple version of the cloud feedback includes: 
	(1) changes in the current resources under the cloud's management (e.g., if some computers crashed); 
	(2) the current resource consumption; 
	(3) termination of jobs; 
	(4) the number of waiting processes of each job, which specifies how many resources the job could use at this moment, if the job were allocated an infinite amount. 

\section{Algorithms}\label{sec:internal}


The internal algorithmic implementation of ERA is encapsulated in separate components -- the \textit{algorithm} and the \textit{prediction} components -- in a flexible ``plug and play'' design, allowing to easily change between different implementations to fit different conditions and system requirements.
The algorithm component is where the actual 
scheduling and pricing of job requests are performed. 
The algorithm may use the prediction component in order to get the market prices or the estimated demand, and the ERA system updates the prediction component online 
with every new request. 

\subsection{Scheduling and Pricing Algorithms} \label{sec:alg}
\paragraph{Interface} 
The ERA algorithm is an online scheduling and pricing algorithm that provides the logic of an ERA system. 
The ERA system forwards queries arriving from users and from the cloud to be answered by the algorithm, and so the internal interface between ERA and the algorithm is similar to the external ERA interface (described in Sections \ref{sec:era-user} and \ref{sec:era-cloud}), 
except that it abstracts 
away all the complexities of interfacing with the external system. 
The main change between these two interfaces is that the algorithm is not given the bids (the monetary value) of the reservation requests, and must decide on the price {\em independently} of the bid. 
It can only make a one-time comparison against the value, 
and the request is accepted as long as the value is not smaller than the price. Thus, the architecture enforces that the algorithm will be monotonic in value (as it sets a threshold price for winning), 
creating an incentive-compatible mechanism with respect to the value; i.e., the 
resulting mechanism is {\em truthful by design}.

The scheduling and pricing of a new job is performed in the \textit{makeReservation} method. As described in detail in Section \ref{sec:era-user}, the input to this method is a reservation request of the form ``I need W cores for T time units, somewhere in the time range [Arrival,Deadline), and will pay at most V for it.'' The answers are of the form ``accept/reject'' and a price P in case of acceptance. The algorithm should also keep track of its planned allocations to actually tell the cloud infrastructure when to run the accepted jobs upon a \textit{getCurrentAllocation} query, and re-plan and optimize upon an \textit{update} query (see Section \ref{sec:era-cloud}).

\paragraph{Basic Econ Scheduling} \label{sec:basic-econ}
The \textit{Basic Econ Scheduling} (Algorithm \ref{alg:basic-econ}) is our basic implementation of an ERA algorithm.
Whenever a new job request 
arrives, the algorithm dynamically sets a price for each time and each unit of the resource (e.g., second*core), 
and the total price is the sum over these unit prices for the requested resources.\footnote{In case of multiple resources, the simple generalization is to set the total price additively over the different types of resources. We choose to focus on additive pricing due to its simplicity and good economic properties (e.g., splitting a request is never beneficial).}
It then schedules the job to start at the cheapest time within its requested window that fits the request, as long as the job's value is not lower than the computed total price.
To determine the price of a resource in a specific time unit $t$ in the future, the algorithm takes into account the amount of resources already promised as well as its prediction for future demand for that time unit. Essentially, the price is set to reflect the externalities imposed on future requests due to accepting this job, according to the predicted demand. The prediction of demand is encapsulated in the prediction component we will discuss in the next section. 

Note that this simple algorithm gives up the flexibility to preempt jobs (swap jobs in and out) and instead allocates to each job a continuous interval of time with a fixed starting time. It also allocates exactly the W requested cores concurrently instead of trading off more time for less parallel cores. We chose to give up these flexibilities in the basic implementation, although they are supported by the ERA API, in order to isolate concerns: this choice separates the algorithmic issues (which are attacked only in a basic way) from pricing issues (which are dealt with) and from learning issues.
In addition, such schedules are robust and applicable under various real-world constraints, and in other cases they may simply be suboptimal and serve as benchmarks.

\begin{algorithm}
\caption{Basic-Econ Scheduling}
\label{alg:basic-econ}
\begin{algorithmic}[1]
\State \textbf{Input}: a new job request \{W*T in [A,D), V\} 
\State \textbf{Output}:  accept or reject, and a price if accepted
\Procedure{Make Reservation}{}
	\For{\textbf{each} $t \in [A,D)$} 	\label{alg:calc-pt-start}
		\State $demand_t() \gets$ the demand estimate function at $t$
		\For{\textbf{each} $i \in [1,W]$}
			\State $price_t(i) \gets$ the highest price $p$ s.t. $demand_t(p)+promised[t]+i > Capacity$
		\EndFor
		\State $cost[t] \gets price_t(1) + price_t(2) + ... + price_t(W)$ 
	\EndFor 														\label{alg:calc-pt-end}
	
	\For{\textbf{each} $t \in [A,D-T]$} \label{alg:calc-totalp-start}
		\State $totalCost[t] \gets cost[t]+ ... + cost[t+T-1]$
	\EndFor 														\label{alg:calc-totalp-end}
	
	\State $t^* \gets \argmin_{t \in [A,D-T]} totalCost[t]$
	\If{$V \geq totalCost[t^*]$}
		\State schedule the job to start at $t^*$ 
		\State \textbf{return} accept at cost $totalCost[t^*]$
	\Else
		\State \textbf{return} reject 
	\EndIf
\EndProcedure
\end{algorithmic}
\end{algorithm}

\subsection{Demand Prediction} \label{sec:prediction}

\paragraph{Interface}

The prediction component is responsible for providing an estimation 
of 
demand at a future time $t$ at any given price, given the current time. Since the inverse function is what we really need, our actual interface provides that inverse function:\footnote{Yet, we present the predictors using both the demand function and its inverse. Moving between the two is straightforward. }
given a future time $t$, the current time, and a quantity of demand $q$, it returns the highest price such that the demand that arrives from the current time till $t$, at this price, is equal to the specified quantity $q$. 

In general, one cannot expect future demand to be determined deterministically -- thus a prediction would, in its most general form, be required to specify a probability distribution over prices that will result in selling the specified quantity. 
As such an object would be hard to work with, our basic implementation simplifies the prediction problem, and requires the predictor to only specify a single price for each demand quantity, as if demand is deterministic. 
Such an approach is justified when the total demand is a result of the aggregation of a large number of independent requests. In that case the demand will be concentrated and the single expected price will reasonably approximate the price distribution. 

\paragraph{Data-based predictors: prediction based on historic data}
ERA's predictor -- the demand oracle -- builds its estimations based on historic data of job requests.
It gets as input a list of past requests, and 
learns, for every time $t$, 
the demand curves (i.e., the demand as a function of price) according to the demand in the list.
Of course, this approach presents multiple challenges: 
first, there is the ``cold start'' problem -- as ERA 
defines a new interface for job requests, there are no past requests of the form that ERA can use to learn. 
Second, the success of the prediction depends on the ability to determine cycles in the demand, such as day-night or days of the week. In addition, the learning methods must also overcome sampling errors and address the non-deterministic nature of the demand (as discussed above). 

Our first implementation of a data-based predictor puts these challenges aside and aims to suggest an approach to address an additional major challenge: the time flexibility of jobs.  
Essentially, the problem is that we expect the predictor to provide the instantaneous demand
, while in ERA the requests are for resources for some duration, within a window of time that is usually longer than the duration. 
Thus, we should answer the following question: how should a job request affect the predicted demand in its requested time window? 



The naive approach would be to spread the demand over the window, e.g., a request of 10 cores for 5 minutes over a window of 50 minutes would contribute 1 core demand in each of the 50 minute window. However this may not reflect the actual demand we should expect. 
For example, consider the input of low-, medium-, and high-value jobs. Each type asks for 100\% of the capacity, where the high-value jobs can run only during the day and the low- and the medium-value jobs can run either day or night. Using the spreading approach we count the demand of the high-value jobs at day, and spread the low- and medium-value jobs over day and night, such that at night we obtain a demand of 50\% of the low- and 50\% of the medium-value jobs. Using this demand gives the impression that we can fill only half of the capacity using the medium-value jobs at night, and so we will set the price to be too low, and will accept low-value jobs at the cost of declining medium-value ones. 

We suggest that this problem can be overcome by taking a different approach based on the LP relaxation of the problem. 
The \textit{LP-based predictor} runs a linear program, offline, to find the optimal (value-maximizing) fractional allocation over past requests, and predicts the demand at time $t$ using the fractional optimal allocation at that time. 
Note that this LP requires many variables -- one variable for every job and every time in the job's time window, and the number of  degrees of freedom may be large, and so one may suspect that the predicted demand will be very different at different times that are experiencing essentially the same demand. 
Our preliminary empirical tests suggest that this LP-based approach is stable, yet future work should test this further and establish theoretical justifications for the approach. 


\section{The ERA System and Simulations}\label{sec:simulator}

ERA is a complete working system: it is implemented as a software package that provides the interfaces described above together with basic implementations of the pricing, scheduling, and prediction algorithms, which are pluggable and can be extended or replaced. 
In addition, the system contains a simulation platform that can simulate the execution of an algorithm given a sequence of job requests and a model of the underlying cloud, using exactly the same core code that is interfaced with the real cloud and users. 
See the system architecture in Figure \ref{fig:architecture1} and a screen-shot of the simulator in Figure \ref{fig:simulator}. 
	
We have performed extensive runs of ERA within the simulator as well as 
proof-of-concept runs with two cloud systems: Hadoop/
YARN and Microsoft's Azure Batch simulator. Next we present a few of these runs to demonstrate the large potential gains of moving from the simple cloud-pricing systems, like the ones currently in use, to ERA -- the Economic Resource Allocation system -- and to demonstrate the ability of the ERA system to integrate with existing cloud systems. 

\paragraph{The importance of economic allocation} 
We first demonstrate the ability of the ERA system to improve the efficiency of cloud resource usage significantly, by 
considering the jobs' values.
We use the simulator with input of jobs that were sampled according to distributions describing a large-scale MapReduce production trace at Yahoo \cite{SWIM}, after some needed modifications of adding deadlines and values that were not included in the original trace. 
In this input, there are 6 classes of jobs, and about 1,400--1,500 jobs of each class. Jobs of class ``yahoo-5'' have the largest average size, and we set them to have a low average value per unit of \$1, while we set jobs of all other classes to have a high value per unit of \$10, to model high-value production jobs. The cluster is way too small to fit all jobs.

We compare ERA's Basic-Econ scheduling algorithm with a gre-edy algorithm that does not take into account the values of the jobs, but instead charges a fixed price per resource unit, and that schedules the job to run at the earliest possible time within its requested time window.
The simulation shows that the greedy algorithm populates most of the cluster with the large, low-value jobs (of class yahoo-5) and results in a low efficiency of only 10\% of the total requested value. In sharp contrast, ERA's Basic-Econ algorithm, which is aware of the values of the jobs and uses dynamic pricing to accept the higher-value jobs, achieves 51\% of the requested value (note that getting 100\% is not possible as the cloud is too small to fit all jobs). 

\paragraph{ERA--Rayon integration}
We next demonstrate that it is feasible to integrate ERA with a real cloud system  
by showing that the cloud succeeds in running real jobs using ERA. In addition, we show that the ERA simulator provides a good approximation to the outcome of the real execution. 
 
We have fully integrated ERA with Rayon \cite{curino2014rayon}, which is a cloud system that handles reservations for computational resources, 
and is part of Hadoop/YARN \cite{yarn} 
(aka MapReduce 2.0). 
The integration required, first, that we introduce economic considerations into the Rayon system, as Rayon's original reservation mechanism did not consider the reservations' monetary valuations. 
Next, we plugged ERA's core code into Rayon's reservation and scheduling process, by adding a layer of simple adapter classes that bridge between ERA's and Rayon's APIs. 
The bridging layer configured Rayon to completely follow ERA's instructions via the {\em getCurrentAllocation} method (see Section \ref{sec:era-cloud}), but made one extension to this query: it added 
an ``empty allocation'' (i.e., allocation of zero resources), for jobs that are during their reservation time-window but are currently not allocated resources. Rayon opened a queue for each job that was returned by ERA, including jobs with an empty allocation, and thus it was able to run jobs earlier than they were scheduled when it was possible. 


We tested the integration by using a workload of MapReduce jobs that we generated using the Gridmix\footnote{http://hadoop.apache.org/docs/r1.2.1/gridmix.html}  platform. The jobs read and wrote synthetic data from files of 100 GB created for this purpose. 
Eight hundred and fifteen jobs were processed, all of which finished successfully. They arrived during a period of one hour, 
asked on average for 3 GB memory, for a duration of 60 seconds on average ($\sigma=6$ seconds). The cluster consisted of 3 nodes, of 80 GB memory each. Rayon's resource manager was configured to use ERA with the simplest greedy algorithm (described above) that allocates a single resource -- GB of memory (as the version of Rayon at the time allocated only memory).


We ran the same job workload in the ERA simulator, with the same greedy algorithm, and a cloud model that communicates with ERA every second with no failures. The comparison between these two runs -- over Rayon (Hadoop) system and in the simulator -- shows that the simulator gives a good approximation to the performance of ERA on a cloud system. 
We found that jobs were scheduled and running on approximately similar points in time and had similar durations. The main difference between the two runs is that while the simulator assigns jobs a constant capacity throughout their (simulated) execution, the real cluster changes their capacity according to various system considerations that are out of ERA's control. The total allocation obtained in these two runs 
(GB*sec) was similar: 76,730 using the simulator vs. 77,056 in the real cloud.


\paragraph{Testing Azure Batch} 

\begin{figure}
\centerline{\includegraphics[scale=0.60]{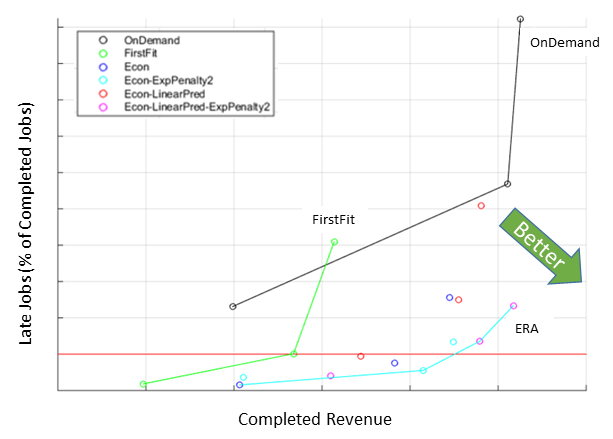}}
\caption{ERA over Azure Batch -- simulation results (axis scales removed). 
ERA's econ algorithm dominates on-demand and first-fit algorithms in terms of the two desired measures of revenue and percentage of late jobs. 
\label{fig:azure_graph} }
\end{figure}

The next set of simulations shows the advantage of using ERA over existing algorithms when applied on a cloud scale.  In a typical cloud environment, we cannot expect one instance of ERA to have complete control of millions of cores. 
Thus, our goal here is to evaluate whether ERA will work with a subset of cores in a region, even while the underlying resource availability is constantly changing. 

The simulations were of a datacenter consisting of 150K cores. ERA was given access to 20\% of the resources and the remaining 80\% 
were allocated to non-ERA requests, 
which were modeled using the standard Azure jobs. 
This means that resources were constantly being allocated/freed in the underlying region and ERA had to account for this.  
The 20\% of the resources under ERA's management came from the pre-emptible resources, but the design does not restrict its use to pre-emptible resources alone.
ERA itself was run as a layer on top of the Azure Batch simulator, which simulates batch workloads on top of the Azure simulator of Microsoft. 

ERA's Basic-Econ scheduling algorithm was experimented relative to two other algorithms:
(1) the on-demand algorithm, which accepts jobs if there are enough available resources to start and run them (availability is checked only for the immediate time, ignoring the duration that the resources are requested). It schedules accepted jobs to run immediately and charges a fixed price; 
(2) the greedy (``FirstFit'') algorithm (described above), which charges the fixed, discounted, price of 65\% of the non-pre-emptible resources price.


A common practice in the industry is to bound the maximal discount over non-pre-emptible machines. Accordingly, in our experiments ERA's Basic-Econ algorithm was restricted so that the price would be no higher than the non-pre-emptible jobs and would give no more than 35\% discount. Several variants of the econ algorithm were explored: 
(1) using either a linear predictor that is based on prior knowledge of the job 
distributions, or a predictor that uses past observations; 
(2) with or without an exponential penalty for later scheduling. Each of the variants was tested at a different capacity of the algorithm's use, so that the higher the capacity the fewer the  resources that remained as spares for re-running failed jobs.

All jobs in the simulation workloads requested a time-window that started at their request-time (i.e., jobs did not reserve in advance). 
As ERA was getting 20\% of the resources, we wanted to evaluate two measure metrics: (1) late-job percentage: this is the percentage of jobs that finished later than their deadlines; (2) accepted revenue: as we can charge only for jobs that are accepted, the better the algorithm, the more jobs we can accept.
Figure \ref{fig:azure_graph} shows that ERA's econ algorithm dominates the other algorithms in terms of these two desired measures.

\section{Grand Challenges}\label{sec:challenges}

Clearly, the main challenge is to get the ERA system integrated in a real cloud system, and 
interface with real paying costumers. 
Short of this grand challenge, there are many research challenges.
In this section we describe several challenges of a practical and theoretical nature related to the ERA (Economic
Resource Allocation) project.

\subsection{Job Scheduling}

There is a vast literature on job scheduling both in the stochastic and adversarial models.
The most obvious related model is job scheduling with laxity, which is the difference between the arrival time and the latest time in which the job can be scheduled and still meet the deadline.
%
The current issues that are raised by our framework give rise to new challenges in both domains.
In our setting it is very reasonable to assume that any job requires only a small fraction of the total resources, and that the
laxity is fairly large compared to the job size. 
An interesting realistic challenge is to have a job give a tradeoff between time (to run) and resources (number of machines), which depends on the degree of parallelism of the job.
Another interesting challenge is to exhibit a model that interpolates between the stochastic model, which gives a complete model of the
job arrival process, and the adversarial model, which does not make any assumptions. It would be nice to have a model that would require only a few parameters and be able to capture many arrival sequences.
Finally, jobs of a reoccurring nature would be very interesting to study  both in the stochastic and adversarial models.

%
%
%

\subsection{Pricing}

In our model we assume that the user has both a clear deadline in mind and an explicit bound on the length of the job.
It would be interesting to give a more flexible guarantee, which
would help the user to set his preferences in a less conservative way. For example, one could allow the job to run after it exhausts
its resources at a certain cost and at a slightly lower priority for a certain additional amount of time. Another similar guarantee is that
the user would have his ``preferred deadline'' and his ``latest deadline'' with a guarantee that most jobs finish at the preferred deadline.
All this is aimed at a more flexible Quality of Service (QoS) guarantee by the system. Pricing such complex guarantees is a significant practical and theoretical challenge.

From the theoretical side, it would be nice to give theoretical guarantees to our system.
First, to show that the users have an incentive to report their information truthfully, and not to try and game the system,
or at least achieve this approximately.
Second, to show that the system reaches a satisfiable steady state (e.g., showing an appropriate equilibrium notion and a related price of anarchy).

%
%

\subsection{Learning}

Our proposed framework requires a significant component of learning. Much of the learning depends on
the observed time series from the past that would be used to predict future requests.
A clear challenge in our setting is to accommodate seasonality effects (daily, such as day versus night; weekly, such as work week versus weekend; annual, such as holidays). Such challenges are well known in the time-series  literature. A more interesting effect is that we have a system where the available resources and the demand are constantly growing, and the challenge is to bundle the two forecasts or somewhat separate them. 
It seems that our prediction model would need a more refined prediction than only the expected value, but for many of our forecasting applications we need to get more detailed information.

An additional uncertainty is that our system might be unable to see certain requests since the user decides that they were unlikely to be accepted and therefore never submitted them. For example, if a more important job is already rejected due to a low value, less-important jobs might be not submitted, and thus the prediction of the demand is even more challenging, given this partial information.

Finally, learning should not be limited only to the forecast of demand, but 
should also 
forecast the accuracy of the requests.
Since in the current system we require that the job will not exceed its maximum length, it is likely to be a conservative estimate, and learning
what is the ``actual'' demand might free significant resources.

\subsection{Robustness}

For any practical system to run it needs a significant level of robustness.
Robustness should take into account both planned and unexpected failures in the various resources.
Modeling this might be done as part of the greater challenge of a QoS guarantee.
We should study what kind of an extreme-case guarantee can we give.
\newpage
{
\small
\bibliographystyle{abbrv}
\bibliography{general}
}
%
%

\end{document}